\documentclass[preprint,aps,showpacs,nofootinbib,preprintnumbers,amsmath,amssymb]{revtex4-1}
\usepackage{amssymb}
\usepackage{epsfig}
\usepackage{graphicx}% Include figure files
\usepackage{subfigure}
\usepackage{dcolumn}% Align table columns on decimal point
\usepackage{bm}% bold math
\usepackage[usenames ,dvipsnames]{xcolor}
\usepackage{slashed}

\begin{document}

\title{CP Violations in Predictive Neutrino Mass Structures}

\author{Chao-Qiang~Geng$^{1,2,3}$\footnote{geng@phys.nthu.edu.tw},
Da~Huang$^{2}$\footnote{dahuang@phys.nthu.edu.tw}
and
Lu-Hsing Tsai$^{2}$\footnote{lhtsai@phys.nthu.edu.tw}}
  \affiliation{$^{1}$Chongqing University of Posts \& Telecommunications, Chongqing, 400065, China\\
  $^{2}$Department of Physics, National Tsing Hua University, Hsinchu, Taiwan\\
  $^{3}$Physics Division, National Center for Theoretical Sciences, Hsinchu, Taiwan
}

\date{\today}
\begin{abstract}
We study the CP violation effects from two types of neutrino mass matrices with (i) $(M_\nu)_{ee}=0$, and (ii) $(M_\nu)_{ee}=(M_\nu)_{e\mu}=0$,
which can be realized by the high dimensional lepton number violating operators $\bar \ell_R^c\gamma^\mu L_L (D_\mu \Phi)\Phi^2$ and $\bar \ell_R^c l_R (D_\mu{\Phi})^2\Phi^2$, respectively.
In (i), the neutrino mass spectrum is in the  normal ordering with the lightest neutrino mass within the range $0.002\,{\rm eV}\lesssim m_0\lesssim 0.007\,{\rm eV}$.
Furthermore, for a given value of $m_0$, there are two solutions for the two Majorana phases $\alpha_{21}$ and $\alpha_{31}$, whereas the Dirac phase $\delta$ is arbitrary.  For (ii), the parameters of $m_0$, $\delta$, $\alpha_{21}$, and $\alpha_{31}$ can be completely determined.
% From the viewpoints of the effect field theory, the high dimensional lepton number violating operators
%$\bar \ell_R^c\gamma^\mu L_L (D_\mu \Phi)\Phi^2$ and $\bar \ell_R^c l_R (D_\mu{\Phi})^2\Phi^2$ can realize the cases (i) and (ii), respectively.
We calculate the CP violating asymmetries in neutrino-antineutrino oscillations for both mass textures of (i) and (ii), which
are closely related to the CP violating Majorana phases.
\end{abstract}

%\pacs{95.35.+d, 98.70.Sa, 13.85.Tp, 14.80.-j}
%\keywords{}
\maketitle
\section{Introduction}
Although it has been established  that neutrinos are massive and mix each other
in the recent several decades~\cite{Anselmann:1992kc,Fukuda:1998mi,Ahmad:2002jz,Ahmad:2002ka,Ahn:2006zza,
Abe:2011sj,An:2012eh},  their nature is still mysterious.
It is known that neutrino mass terms could be of the Dirac type, in analogy to the charged fermions, $i.e.$ quarks and charged leptons, or  the Majorana type, possibly generated by the Weinberg operator $\bar L^c L\Phi\Phi$~\cite{Weinberg:1980bf}. In the literature, there have been
 many models to realize the Weinberg operator at the tree~\cite{TypeIseesaw1,TypeIseesaw2,TypeIseesaw3,TypeIseesaw4,TypeIseesaw5,typeIIseesaw1,typeIIseesaw2,typeIIseesaw3,
typeIIseesaw4,typeIIseesaw5,typeIIseesaw6,typeIIseesaw7,Foot:1988aq} and loop~\cite{Krauss:2002px,Ma:2006km,Aoki:2008av,Gustafsson:2012vj,Zee:1980ai,Zee:1985id,Babu:1988ki} levels. Note that the current neutrino oscillation experiments cannot determine the three CP violating phases, especially for the two Majorana phases, which is an important problem in neutrino physics.

The Weinberg operator violates the lepton number symmetry by two units, but sometimes it is not the one that gives the dominant contribution to the Majorana neutrino masses or the lepton number violating (LNV) processes. Instead, other higher dimensional LNV operators, for examples, the dimension-7,  ${\cal O}_7 = \bar \ell_R^c \gamma_\mu L_L (D_\mu\Phi)\Phi^2$~\cite{delAguila:2012nu, delAguila:2013zba,Aparici:2013xga,Geng:2015qha}, and  dimension-9,  ${\cal O}_9 = \bar \ell_R^c\ell_R(D\Phi)^2\Phi^2$~\cite{Chen:2006vn,Chen:2007dc,delAguila:2011gr,Gustafsson:2014vpa, King:2014uha,Geng:2014gua}, operators can lead to new Majorana neutrino mass structures different from those by the Weinberg operator if they are prominent. Specifically, due to the nontrivial dependence of the charged lepton masses, ${\cal O}_7$ generically generates a neutrino mass matrix with $(M_\nu)_{ee}=0$~\cite{Geng:2015qha} in the flavor basis, while ${\cal O}_9$ naturally gives rise to the texture with $(M_\nu)_{ee}=(M_\nu)_{e\mu}=0$~\cite{Chen:2006vn,delAguila:2011gr,Gustafsson:2012vj, delAguila:2012nu,Gustafsson:2014vpa,Geng:2014gua}.
By fitting the present neutrino oscillation data, both textures predict that the neutrino mass matrix should be of the normal ordering, and already give stringent constraints to the unknown parameters in the neutrino mass matrix. In particular, in Refs.~\cite{Geng:2014gua,Geng:2015qha}, we show
 that they could naturally lead to nontrivial values for the three CP violating phases. We regard that the higher dimensional operators would provide us
 with a new way to generate the new neutrino structures, besides the ordinary approach by imposing flavor symmetries~\cite{Ma:2001dn,Babu:2002dz,Ma:2005pd,Lam:2008sh,Petcov:2014laa,Ma:2015gka, Girardi:2015vha,Ma:2015pma,Acosta:2014dqa, Ballett:2014dua, Ballett:2013wya}.

In the present paper, we investigate a relevant question: to what extent can the conditions $(M_\nu)_{ee}=0$ and $(M_\nu)_{ee}=(M_\nu)_{e\mu}=0$ restrict the neutrino mass matrix structure, especially the leptonic CP violating phases, based on measured quantities from oscillation experiments? In our treatment, we also take into account the experimental uncertainties in the data in order to see their effects on the results. Note that there were already many studies about
these two specific neutrino mass matrices in the literature, see {\it e.g.}, Refs.~\cite{Frampton:2002yf,Xing:2002ta,Xing:2002ap,Desai:2002sz,Guo:2002ei,Honda:2003pg, Merle:2006du,Dev:2006qe,Kumar:2011vf,Fritzsch:2011qv,Ludl:2011vv,Meloni:2012sx,Grimus:2012zm, Grimus:2012ii,Cebola:2015dwa,Gautam:2015kya,Dev:2015lya,Acosta:2012qf}, in which more texture-zero neutrino mass matrices were examined. Here, our focus  is their implications on the leptonic CP violating phases.

With the predicted neutrino mass parameters, the next question is how to test the above two texture-zero structures by measuring all the relevant parameters in the neutrino mass matrices, especially the non-trivial Majorana phases. Previous studies showed that neutrino-antineutrino oscillations gave us
{prospective}
%limited
%one of the only few
approaches to probe the Majorana phases~\cite{Kayser:1984ge,Langacker:1998pv,deGouvea:2002gf,Xing:2013ty,Xing:2013woa, Zhou:2013eoa,Xing:2014yka,Xing:2014eia,Xing:2015zha,Wang:2015rma}, which is impossible for the conventional (anti)neutrino-(anti)neutrino oscillation experiments. We find that once the possible regions of these phases are depicted for the present two textures,
%%%%%%%%%%%?????????
%the probability and
the associated CP violating asymmetries of the neutrino-antineutrino oscillations can be predicted. As will be shown later, by appropriately choosing the (anti)neutrino beam energy and baseline length, some of the asymmetries can be of  $O(1)$.

This paper is organized as follows. In Sec.~II we study the implications of the texture-zero conditions $(M_\nu)_{ee}=0$ and $(M_\nu)_{ee}=(M_\nu)_{e\mu}=0$ to the unknown neutrino mass parameters, including the lightest neutrino mass and the three CP violating phases, based on the existing data. With the preferred values of these parameters, we predict the CP violating asymmetries in the neutrino-antineutrino oscillations for both textures in Sec.~III. In Sec.~IV, we give a short summary.

\section{Texture-Zero Neutrino Mass Matrix}
As the Majorana neutrino mass matrix $M_\nu$ is symmetric,  there are six independent complex elements $(M_{\nu})_{ee}$, $(M_{\nu})_{e\mu}$, $(M_{\nu})_{e\tau}$, $(M_{\nu})_{\mu\mu}$, $(M_{\nu})_{\mu\tau}$, and $(M_{\nu})_{\tau\tau}$. A well-defined $M_\nu$ can be connected with the observed quantities from neutrino oscillations. Up to the field redefinition, all of the above matrix elements depend on the nine neutrino parameters, including 3 masses, 3 mixing angles, 1 Dirac CP phase and 2 Majorana CP phases. In the flavor basis, where the charged lepton mass matrix is diagonal, the
neutrino mass matrix defined in the Lagrangian $L=-{1\over 2}\overline{(\nu_{L\ell})^c} M_{\ell\ell'}\nu_{L\ell'}+{\rm H.c.}$ can be decomposed as follows,
\begin{eqnarray}
M_\nu\equiv V^*{\rm diag}(m_1,m_2,m_3) V^\dagger\;,\label{Eq_Mn}
\end{eqnarray}
where $m_{1,2,3}$ are three neutrino masses. $V$ is the charged current leptonic mixing matrix~\cite{Pontecorvo:1957cp,Maki:1962mu}, conventionally expressed in the standard parametrization as~\cite{Chau:1984fp,Agashe:2014kda}
\begin{eqnarray}
V=\left(\begin{array}{ccc}
c_{12}c_{13}&s_{12}c_{13}&s_{13}e^{-i\delta}\\
-s_{12}c_{23}-c_{12}s_{23}s_{13}e^{i\delta}&c_{12}c_{23}-s_{12}s_{23}s_{13}e^{i\delta}&s_{23}c_{13}\\
s_{12}s_{23}-c_{12}c_{23}s_{13}e^{i\delta}&-c_{12}s_{23}-s_{12}c_{23}s_{13}e^{i\delta}&c_{23}c_{13}\\
\end{array}\right)
\left(\begin{array}{ccc}
1&0&0\\
0&e^{i\alpha_{21}/2}&0\\
0&0&e^{i\alpha_{31}/2}\\
\end{array}
\right)\,,~\label{Eq_Vlepton}
\end{eqnarray}
where $s_{ij}\equiv \sin \theta_{ij}$, $c_{ij}\equiv \cos \theta_{ij}$, $\delta$ is the Dirac phase, and $\alpha_{21,31}$ represent two Majorana phases within the range $[0,\,2\pi]$. The values of $\theta_{12}$, $\theta_{23}$, $\theta_{13}$, $\Delta m_{21}^2$ and $|\Delta m_{32}^2|$ have already been obtained from the neutrino oscillation experiments~\cite{Agashe:2014kda}, so that the rest four unknown neutrino parameters are the three CP phases, $\delta$, $\alpha_{21}$
and $\alpha_{32}$, and the lightest neutrino mass, $m_0$. Note that only the absolute value $\Delta m_{32}^2$ has been acquired, which leaves us two possible orderings: the normal ordering for $\Delta m_{32}^2 >0$ with $(m_1,m_2,m_3)=(m_0,\sqrt{m_0^2+\Delta m_{21}^2},\sqrt{m_0^2+\Delta m_{31}^2})$, and the inverted one for $\Delta m_{32}^2 < 0$ with $(m_1,m_2,m_3)=(\sqrt{m_0^2-\Delta m_{31}^2},\sqrt{m_0^2-\Delta m_{31}^2+\Delta m_{21}^2},m_0)$.

$M_\nu$ can have some special approximate texture-zero forms when it is generated by some  high dimensional LNV operators.
For example, if $\mathcal{O}_7=\bar{\ell^c_R} \gamma^\mu L_L (D_\mu\Phi)\Phi^2$ gives the leading contribution to neutrino masses, then $(M_\nu)_{\ell\ell'}$ should be approximately proportional to the sum of charged lepton masses, $m_\ell+m_{\ell'}$, with $\ell$ and $\ell'=e,\,\mu,\,\tau$. Consequently, $(M_{\nu})_{ee}$ should be much smaller than other elements. Similarly, if $\mathcal{O}_9 = \bar{\ell^c_R} \ell_R (D_\mu \Phi)^2 \Phi^2$ dominates over other LNV operators, $(M_\nu)_{\ell\ell'}$ will be proportional to $m_\ell m_{\ell'}$. It turns out that not only $(M_\nu)_{ee}$ but also $(M_\nu)_{e\mu}$ are expected be greatly suppressed due to the hierarchy in the charged lepton masses. In other words, the neutrino mass matrices obtained from these LNV effective operators are characterized by the special zero textures $(M_\nu)_{ee}=0$ and 
$(M_\nu)_{ee}=(M_\nu)_{e\mu}=0$\footnote{Besides  the relative smallness of the element $(M_\nu)_{ee}$ already argued in the main text 
for the two high-dimensional effective operators of ${\cal O}_7$ and ${\cal O}_9$, its absolute value is further constrained by the neutrinoless double beta ($0\nu\beta\beta$) decay processes~\cite{Agostini:2013mzu,KamLANDZen,Gando:2012zm,Argyriades:2008pr, Arnaboldi:2008ds,Arnold:2005rz,Barabash:2010bd}. Note that these two effective operators give the dominant contributions to the $0\nu\beta\beta$ decay at tree level, while the Majorana mass terms arising from ${\cal O}_{7(9)}$ begins at one-(two-)loop level. Due to the absence of the loop suppression, these two operators are more sensitive to the $0\nu\beta\beta$ decay processes, which constrain the cutoff scales and Wilson coefficients of the  effective operators greatly and lead to the conclusion that $(M_\nu)_{ee} < 10^{-13}$~eV.  For further details, please refer to Refs.~\cite{Geng:2015qha} and \cite{Geng:2014gua}.}, 
the implications of which will be discussed in detail in the following two subsections.

\subsection{$(M_\nu)_{ee}=0$}
\begin{figure}
\includegraphics[width=16cm]{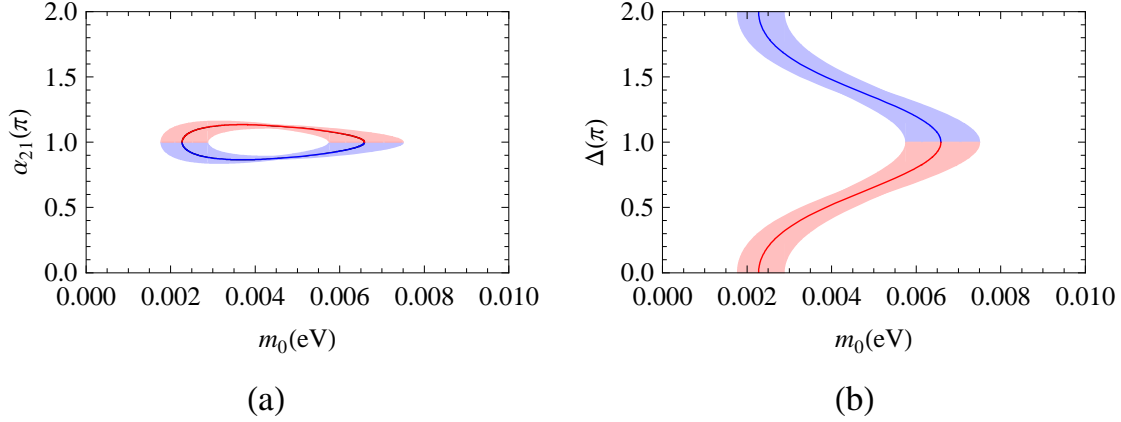}
\caption{Contours for $(M_{\nu})_{ee}=0$ in  (a) $m_0$-$\alpha_{21}$ and (b) $m_0$-$\Delta$ planes, respectively. Solid lines with experimental central values and $1\sigma$ standard deviation, where the colors of red and blue indicate that the values of $\sin\alpha_{21}$ are positive and negative, respectively.}
\label{Fig_Mee0}
\end{figure}

By expanding the right-hand side of Eq.~(\ref{Eq_Mn}) with the standard parametrization of $V$ in Eq.~(\ref{Eq_Vlepton}), the condition $(M_\nu)_{ee}=0$ can be transformed into the following relation,
\begin{eqnarray}
(M_\nu)_{ee}&=&c_{12}^2 c_{13}^2 m_1+s_{12}^2c_{13}^2m_2 e^{-i\alpha_{21}}+s_{13}^2m_3 e^{-i\Delta}=0\,,\label{Eq_Mee}
\end{eqnarray}
where the phase $\Delta\equiv \alpha_{31}-2\delta$ is defined, which will be used to replace $\alpha_{31}$ as an independent Majorana phase hereafter.
Note that this equation excludes the inverted ordering at more than $2\sigma$ significance by current oscillation experiment results~\cite{Agashe:2014kda},
so that we only need to consider the normal-ordering neutrino mass matrix from now on. Fig.~\ref{Fig_Mee0} shows the allowed parameter space region
satisfying Eq.~(\ref{Eq_Mee}), in which the solid curves represent the parameters when the experimental observables are at their central values
in PDG~\cite{Agashe:2014kda}, while the shadow areas correspond to the 1$\sigma$ standard deviations.
It is interesting to note that $m_0$ and $\alpha_{21}$ are already limited within small parameter regions, with $0.8\pi\lesssim \alpha_{21}\lesssim 1.2\pi$ and $0.0015\, {\rm eV}\, \lesssim m_0 \lesssim 0.008 \,$eV, respectively. In particular, the extremal values of the lightest neutrino mass $(m_0)_{\rm min (max)}$ are related to two CP conserving solutions to Eq.~(\ref{Eq_Mee}) with $\alpha_{21} = \pi$ and $\Delta=0(\pi)$,
\begin{eqnarray}
s_{12}^2c_{13}^2m_2-c_{12}^2c_{13}^2m_1-s_{13}^2m_3=0\;\;\; {\rm and}\;\;
c_{12}^2c_{13}^2m_1-s_{12}^2c_{13}^2m_2-s_{13}^2m_3=0\,.
\end{eqnarray}
For each value of $m_0$ within the regions $[(m_0)_{\rm min},\, (m_0)_{\rm max}]$, there exist two solutions for $\alpha_{21}$ and $\Delta$, differentiated by the positive or negative $\sin\alpha_{21}$, which are shown in Fig.~\ref{Fig_Mee0} as red or blue curves/shadows. 
Another interesting observation is that the obtained $\alpha_{21}$ is limited around $\pi$, which can be understood directly from Eq.~(\ref{Eq_Mee}). Since $s_{13}^2$ is very small, the third term in Eq.~(\ref{Eq_Mee}) can be neglected, and the first two terms must balance each other to achieve the constraint of the vanishing $(M_\nu)_{ee}$, which only requires $\alpha_{21} \sim \pi$ in order to reverse the sign of the second term. Moreover,
$\alpha_{21}$ is precisely predicted to be $1.1\pi$ or $0.9\pi$ when $m_0$ is located within $0.004\,{\rm eV}\lesssim m_0\lesssim0.005\,{\rm eV}$\footnote{Similar results are also given in Ref.~\cite{Xing:2015zha}.}, no matter how experimental errors vary. Finally, we remark that $(M_\nu)_{ee}=0$ does not provide any constraint on $\delta$, which is only contained in $\Delta$. If one focus on the real $M_\nu$, then $\delta$ can be taken as $0$ or $\pi$, for the cases $m_0=(m_0)_{\rm min}$ and $(m_0)_{\rm max}$. Therefore, there are 4 independent real neutrino mass matrices for $(M_\nu)_{ee}=0$.

\subsection{$(M_\nu)_{ee}=(M_\nu)_{e\mu}=0$}
In this subsection, we will concentrate on the case with $(M_\nu)_{ee}=(M_\nu)_{e\mu}=0$. Note that such constraints correspond to two complex equations, which enable us to uniquely solve for the remaining four parameters $(m_0,\delta,\alpha_{21},\Delta)$ in the neutrino mass matrix undetermined from the current oscillation experiments. Now we sketch the procedure of deriving these quantities in terms of the measured observables~\cite{Gustafsson:2012vj}. The first step is to write down the two conditions in the parametrization independent form:
\begin{eqnarray}
(M_\nu)_{ee}&=&m_1V_{e1}^{*2}+m_2V_{e2}^{*2}+m_3V_{e3}^{*2}=0\,,\nonumber\\
(M_\nu)_{e\mu}&=&m_1V_{e1}^*V_{\mu1}^*+m_2V_{e2}^*V_{\mu2}^*+m_3V_{e3}^*V_{\mu3}^*=0\,,
\end{eqnarray}
with which we can obtain the following useful formulas
\begin{eqnarray}
&&1-{\Delta m_{21}^2\over \Delta m_{31}^2}={1-|Y|^2\over 1-|X|^2}\,,\label{Eq_deltaDerived}\\
&&{m_2\over m_3}={|Y|\over |X|}\,,\label{Eq_m0Derived}\\
&&{\rm Im}(V_{e3}V_{\tau3}V_{e2}^*V_{\tau2}^*)={\rm Im}(V_{e1}V_{\tau1}V_{e3}^*V_{\tau3}^*)
={\rm Im}(V_{e2}V_{\tau2}V_{e1}^*V_{\tau1}^*)=0\label{Eq_alphaDerived}\,,
\end{eqnarray}
with
\begin{eqnarray}
X={V_{e2}V_{\tau2}V_{e1}^*V_{\tau 1}^*\over |V_{e1}|^2|V_{e1}V_{\mu3}-V_{e3}V_{\mu1}|^2}\,,\,
Y^{-1}={V_{e1}V_{\tau1}V_{e3}^*V_{\tau 3}^*\over |V_{e3}|^2|V_{e3}V_{\mu2}-V_{e2}V_{\mu3}|^2}\,.
\end{eqnarray}
Since neither $|X|$ nor $|Y|$ depends on the two Majorana phases and $m_0$, we can determine the Dirac phase $\delta$ from Eq.~(\ref{Eq_deltaDerived}). By substituting the obtained Dirac phase  into Eq.~(\ref{Eq_m0Derived}), we can solve for $m_0$. Finally, two Majorana phases can be fixed with Eq.~(\ref{Eq_alphaDerived}). In the standard parametrization, the solution is expressed by~\cite{Gustafsson:2012vj}
\begin{subequations}\label{Eqn_Soln}
\begin{eqnarray}
\cos\delta&=&{s_{13}^{-1}\over{2(1+t_{12}^2)+r(1-t_{12}^2)}}\Big\{{t_{12}\over t_{23}}r(t_{13}^{-2}-1)
-{t_{23}\over t_{12}}\Big[(1-t_{12}^4)+{r\over2}(1+t_{12}^4)\Big]\Big\}\,,\\
\Delta&=&\arg(-s_{13}+t_{12}t_{23}e^{-i\delta})\;,\\
\alpha_{21}&=&\arg\Big({s_{13}-t_{12}t_{23}e^{-i\delta}\over s_{13}+t_{12}^{-1}t_{23}e^{-i\delta}}\Big)\,,\\
m_0&=&\sqrt{\Delta m_{21}^2{(2+r)\over 2r}{t_{13}^2(t_{12}^2t_{23}^2-2s_{13}t_{12}t_{23}c_\delta+s_{13}^2)\over 1-t_{13}^2(1+t_{12}^2t_{23}^2-2s_{13}t_{12}t_{23}c_\delta)}}\,,
\end{eqnarray}
\end{subequations}
with $r\equiv \Delta m_{21}^2/(\Delta m_{32}^2+\Delta m_{21}^2/2)$.
Note that for each value of $m_0$, we can obtain two solutions of the CP violating phases ($\delta$, $\alpha_{21}$, and $\Delta$), which can be connected with each other by the replacements of  $\delta\rightarrow 2\pi-\delta$, $\alpha_{21}\rightarrow 2\pi-\alpha_{21}$, and $\Delta\rightarrow 2\pi-\Delta$.

\begin{figure}
\includegraphics[width=16cm]{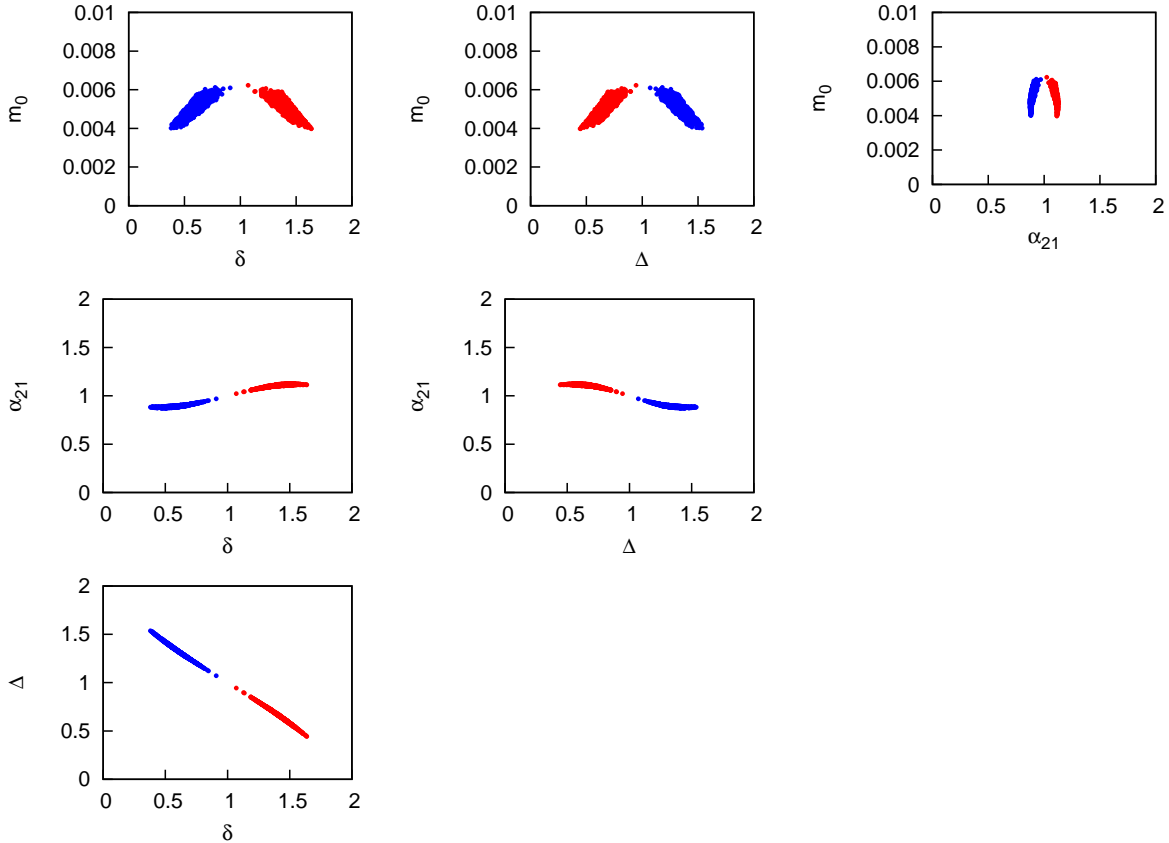}
\caption{Correlations among the parameters: $\delta$, $\Delta$, $\alpha_{21}$, and $m_0$, where the red (blue) color represents that $\sin\alpha_{21}>0\,(<0)$.}
% are positive and negative, respectively.}
\label{Fig_params}
\end{figure}

Fig.~\ref{Fig_params} shows the allowed parameter space according to Eq.~(\ref{Eqn_Soln}) when we take the fitting values of the five parameters ($\theta_{12},\theta_{23},\theta_{13},\Delta m_{21}^2,\Delta m_{32}$) within the $1\sigma$ deviation as the input parameters, in which the red (blue) areas label the regions with $\sin\alpha_{21} >0(<0)$. If we take the central experimental values of the measured quantities~\cite{Agashe:2014kda}, two solutions can be obtained with $m_0=5.07\times10^{-3}\,{\rm eV}$, $\delta=0.59\pi\;(1.41\pi)$, $\alpha_{21}=0.89\pi\;(1.11\pi)$, and $\Delta=1.34\pi\;(0.66\pi)$. The corresponding leptonic Jarlskog invariant, $J=c_{12}c_{13}^2 c_{23} s_{12} s_{13}s_{23}s_\delta$, is equal to 0.033($-0.033$), which characterizes
CP violation in the lepton sector. It is worth {noting} that the value of $\delta=1.41\pi$ for one of the solution is close to the central value of $\delta$ from the global fitting result~\cite{Agashe:2014kda}. Moreover, in the present texture-zero case, when Dirac phase $\delta$ is taken to be the CP conserving values,
such as $\delta=0$ or $\pi$, the two Majorana ones can only be CP conserving values too, {\it i.e.}, $\alpha_{21}$ and $\Delta$ should be 0 or $\pi$.
However, with the results shown in Fig.~\ref{Fig_params}, it is interesting that the CP conserving cases are excluded at the 1$\sigma$ level. Finally, if the Dirac phase is taken to be of the maximal CP violating value with $\delta = \pi/2$ ($-\pi/2$), the Majorana phases are predicted to be $\alpha_{21}=0.88\pi\;(1.12\pi)$ and $\Delta=1.42\pi\;(0.58\pi)$ with the experimental central values for the mixing angles.

%\*************************************************

%\begin{figure}
%\includegraphics[width=16cm]{invariantQ.eps}
%\caption{Correlations among the rephasing invariant quantities for NO.}
%\label{Fig_invariantQ}
%\end{figure}

\section{Neutrino-Antineutrino Oscillations}
With the non-trivial Majorana CP violating phases, especially for the case with $M_{ee}=M_{e\mu}=0$, the immediate important question is how to measure them. Traditional (anti)neutrino-(anti)neutrino oscillation experiments can be used to measure the Dirac phase, but they are insensitive to the Majorana ones since the involved  processes are lepton number conserving so that the Majorana phases are cancelled out in the corresponding formulas. As a result, in order to measure the Majorana phases, one of the necessary conditions is that the involved processes are LNV. As pointed out in Refs.~\cite{Kayser:1984ge,Langacker:1998pv,deGouvea:2002gf,Xing:2013ty}, the neutrino-antineutrino oscillations provide us with a promising way to detect them. Unfortunately, the neutrino-antineutrino channels suffer from an additional helicity suppression factor $(m_\nu/E)^2$ in the oscillation probabilities, compared with the corresponding usual (anti)neutrino-(anti)neutrino oscillation channels. Therefore, it is challenging to carry out such experiments.
The use of  the low-energy M$\ddot{\rm o}$ssbauer electron antineutrinos~\cite{Mossbauer} with 18.6 keV, which are emitted from the bound-state beta decay of $^3$H to $^3$He,  can improve the situation greatly by enhancing the signal by a factor of ${\cal O}(10^4)$ as compared with the conventional reactor antineutrinos~\cite{Xing:2013ty}. However, even in this case, it is still practically impossible to observe these neutrino-antineutrino oscillations, as will be shown below.

The general formulas for the neutrino-antineutrino oscillation probabilities $P(\nu_\alpha \to \bar{\nu}_\beta)$ and $P(\nu_\beta \to \bar{\nu}_\alpha)$ in the three-flavor framework are~\cite{Xing:2013ty}
\begin{eqnarray}
P(\nu_\alpha\rightarrow \bar\nu_\beta)&=&{|K|^2\over E^2}\Big[|M_{\alpha\beta}|^2-4\sum_{i<j}m_i m_j
\mathrm{Re}(V_{\alpha i}V_{\beta i}V_{\alpha j}^*V_{\beta j}^*)\sin^2\Big({\Delta m_{ji}^2L\over 4E}\Big)\nonumber\\
 &&+2\sum_{i<j}m_i m_j
\mathrm{Im}(V_{\alpha i}V_{\beta i}V_{\alpha j}^*V_{\beta j}^*)\sin\Big({\Delta m_{ji}^2L\over 2E}\Big)\Big]\;,\\
P(\bar\nu_\alpha\rightarrow \nu_\beta)&=&{|\bar K|^2\over E^2}\Big[|M_{\alpha\beta}|^2-4\sum_{i<j}m_i m_j
\mathrm{Re}(V_{\alpha i}V_{\beta i}V_{\alpha j}^*V_{\beta j}^*)\sin^2\Big({\Delta m_{ji}^2L\over 4E}\Big)\nonumber\\
 &&-2\sum_{i<j}m_i m_j
\mathrm{Im}(V_{\alpha i}V_{\beta i}V_{\alpha j}^*V_{\beta j}^*)\sin\Big({\Delta m_{ji}^2L\over 2E}\Big)\Big]\;,
\end{eqnarray}
where $K$ and $\bar K$ are the kinetic factors with $|K|=|\bar K|$ and $L$ is the neutrino traveling length. 
 Now it is interesting to estimate the neutrino-antineutrino oscillation probabilities for different channels to see if they have the potential to be observed under the present experimental status, especially the M$\ddot{\rm o}$ssbauer neutrinos advertised in Ref.~\cite{Xing:2013ty}. By assuming the kinematic factor $K\sim {\cal O}(1)$, electron antineutrino energy  $E \sim 18.6$~keV, and  oscillation baseline length  $L \sim 300$~m, we can obtain the largest $\nu_e - \bar{\nu}_e$ oscillation probability to be $P(\nu_e \to \bar{\nu}_e) \sim  {\cal O}(10^{-13})$ for $m_0 = 0.0065$~eV. The largest probabilities for other oscillation channels,
such as $P(\bar{\nu}_e \to \nu_\mu )$, would be of the similar order. In the view of these simple exercises, it seems impossible to observe these oscillations practically in the foreseeable experiments. 

It is obvious that $P(\nu_\alpha\rightarrow \bar\nu_\beta)$ and its CP conjugate process $P(\nu_\alpha\rightarrow \bar\nu_\beta)$ can have different values when $V$ is complex, which is the origin of  CP violation in the lepton sector. Therefore, we can define the CP asymmetry parameter $A_{\alpha\beta}$ by
%to characterize it.
\begin{eqnarray}\label{Eqn_asymmetry}
A_{\alpha\beta}&\equiv&{P(\nu_\alpha\rightarrow \bar\nu_\beta)-P(\bar\nu_\alpha\rightarrow \nu_\beta)\over P(\nu_\alpha\rightarrow \bar\nu_\beta)+P(\bar\nu_\alpha\rightarrow \nu_\beta)}\nonumber\\
&=&{2\sum_{i<j}m_i m_j
\mathrm{Im}(V_{\alpha i}V_{\beta i}V_{\alpha j}^*V_{\beta j}^*)\sin\big((f\pi\Delta m_{ji}^2)/(2\Delta m_{21}^2)\big)\over |M_{\alpha\beta}|^2-4\sum_{i<j}m_i m_j
\mathrm{Re}(V_{\alpha i}V_{\beta i}V_{\alpha j}^*V_{\beta j}^*)\sin^2\big((f\pi\Delta m_{ji}^2)/(4\Delta m_{21}^2)\big)}\,,
\end{eqnarray}
where $f=(L/E)(\Delta m_{21}^2/\pi)$.
In the following, we shall use the obtained CP violating phases from the previous two texture-zero neutrino mass matrices to predict the oscillation probabilities and the associated CP violating asymmetries in some neutrino-antineutrino oscillation channels of great phenomenological interest, and then see how these measurements can help us to probe or constrain the whole picture of neutrino masses.

\subsection{$(M_\nu)_{ee}=0$}
\begin{figure}
\includegraphics[width=16cm]{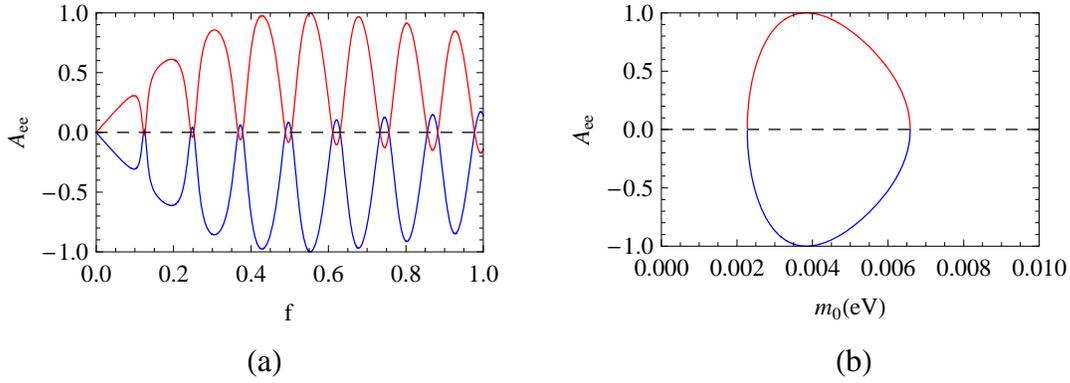}
\caption{$A_{ee}$ as functions of (a) $m_0$ and (b) $f$, where the red (blue) corresponds to $\sin\alpha_{21}>0\, (<0)$.}
% are positive and negative, respectively.}
\label{Fig_Aee}
\end{figure}

According to Eq.~(\ref{Eqn_asymmetry}), the asymmetry $A_{ee}$ can be expressed by
\begin{eqnarray}
A_{ee}&=&2\Big[m_1m_2c_{12}^2c_{13}^4s_{12}^2\sin^2{f\pi\over 2}\sin{\alpha_{21}}
+m_1m_3 c_{12}^2c_{13}^2s_{13}^2\sin^2\Big[{f\pi \over 2} \Big(1+{\Delta m_{32}^2\over\Delta m_{21}^2}\Big)\Big]\sin\Delta\nonumber\\
&&-m_2m_3c_{13}^2s_{12}^2s_{13}^2\sin^2{f\pi\Delta  m_{32}^2\over 2\Delta  m_{21}^2}\sin(\alpha_{21}-\Delta)\Big]
\Big/\Big[ -|(M_\nu)_{ee}|^2\nonumber\\
&&+4m_1m_2c_{12}^2c_{13}^4s_{12}^2\sin^2{f\pi\over 4}\cos\alpha_{21}
+4m_1m_3 c_{12}^2c_{13}^2s_{13}^2\sin^2\Big[{f\pi \over 4} \Big(1+{\Delta m_{32}^2\over\Delta m_{21}^2}\Big)\Big]\cos\Delta\nonumber\\
&&+4m_2m_3c_{13}^2s_{12}^2s_{13}^2\sin^2{f\pi\Delta  m_{32}^2\over 4\Delta  m_{21}^2}\cos(\alpha_{21}-\Delta)\Big]\,.
\end{eqnarray}
%where the value of $A_{ee}$ is proportional to the linear combination of $\sin\alpha_{21}$ and $\sin\Delta$.
When imposing the condition $(M_\nu)_{ee}=0$, both $\alpha_{21}$ and $\Delta$ can be expressed as the functions of $m_0$.
Therefore, by fixing the factor $f$ to some definite value, the CP violating asymmetries of various channels can also have definite values for every $m_0$.
As an illustration,
Fig.~\ref{Fig_Aee}a gives the correlation between $A_{ee}$ and $m_0$ when $f=0.55$, where we only take the central values of the measured quantities in our calculation. By comparing Fig.~\ref{Fig_Aee}a and Fig.~\ref{Fig_Mee0}a, we see that the detection of $A_{ee}$ directly implies the existence of a nonzero $\sin\alpha_{21}$, and their signs are positively correlated. Our direct calculation confirms this observation.
We also plot the variation of $A_{ee}$ against the factor $f$ in Fig.~\ref{Fig_Aee}b by taking $m_0=0.004$~eV, which shows that if we can fine tune the beam energy $E$ or the baseline length $L$ to make an appropriate value of $f$, a large CP violating asymmetry in the $\nu_e$-$\bar{\nu}_e$ channel $|A_{ee}|\simeq 1$ can be obtained. Furthermore, note that the dependence of $A_{ee}$ on the Dirac phase $\delta$ is only through the combination of $\Delta = \alpha_{31}-2\delta$, so that even if $\delta$ vanishes, there can still be quite sizable CP-violating effects in the $\nu_e$-$\bar{\nu}_e$ oscillation experiment due to the compensation from $\alpha_{31}$ in $\Delta$, which starkly shows the significance of the Majorana phases in generating the CP-violating effects.
 Finally, we make the estimation of the asymmetries in other neutrino-antineutrino oscillation channels, by simply taking $f=0.55$, $m_0=0.004$~eV, and $\delta=0$ or $\pi/2$, together with other measured observables at their central values, with the results shown in Table~ \ref{Tab_Asymmetry}.
\begin{table}[ht]
\caption{Asymmetries for neutrino-antineutrino oscillations with
%$\delta=0$ and $\pi/2$, where we have set
$f=0.55$ and $m_0=0.004\,{\rm eV}$.}
\begin{tabular}{ccccccc}
\hline
$\delta$&$A_{ee}$&$A_{e\mu}$&$A_{e\tau}$&$A_{\mu\mu}$&$A_{\mu\tau}$&$A_{\tau\tau}$\\
\hline
0&1.0(-1.0)&0.38(-0.38)&-0.71(0.71)&0.19(-0.19)&-0.21(0.21)&0.17(-0.17)\\
$\pi/2$&1.0(-1.0)&0.61(0.20)&-0.20(-0.47)&-0.26(0.25)&0.18(-0.19)&-0.15(0.16)\\
\hline
\end{tabular}
\label{Tab_Asymmetry}
\end{table}

\subsection{$(M_\nu)_{ee}=(M_\nu)_{e\mu}=0$}
It was shown previously that all of the mass and mixing parameters can be fixed when $(M_{\nu})_{ee}=(M_{\nu})_{e\mu}=0$. It follows that all of the CP violating asymmetries in the neutrino-antineutrino oscillations can also be determined. We remark that for this case $A_{e\tau}$ always vanishes since each term in the summation of Eq.~(\ref{Eqn_asymmetry}) is zero as the consequence of Eq.~(\ref{Eq_alphaDerived}).
Using the central values of neutrino mixing parameters from neutrino oscillations with $f=3.5$, we can predict $A_{ee}=0.92$, $A_{e\mu}=-0.09$, $A_{\mu\mu}=-0.15$, $A_{\mu\tau}=0.1$, and $A_{\tau\tau}=-0.09$. Therefore, the $\nu_e$-$\bar{\nu}_e$ oscillation is the most prospective channel to probe this neutrino mass texture.

\section{Summary}
%Conclusion}
We have studied the CP violating asymmetries and related LNV processes such as the neutrino-antineutrino oscillations under two types of the neutrino mass textures, $(M_\nu)_{ee}=0$ and $(M_\nu)_{ee}=(M_\nu)_{e\mu}=0$, realized by
the high dimensional lepton number violating operators. For $(M_\nu)_{ee}=0$, there are two solutions of $\alpha_{21}$ and $\Delta$ for each value of $m_0$, with $0.002\,{\rm eV}\lesssim m_0 \lesssim 0.007\,{\rm eV}$ and an arbitrary value of $\delta$.
For $(M_\nu)_{ee}=(M_\nu)_{e\mu}=0$, two solutions for free parameters ($m_0$, $\delta$, $\alpha_{21}$, $\Delta$) can be obtained,
in which one of them with $\delta=1.41\pi$ is close to the global fitting result.
The effect of the nonzero values of the two Majorana phases can be reflected by the related CP violating asymmetry parameters $A_{\alpha\beta}$ in neutrino-antineutrino oscillations. In the texture $(M_\nu)_{ee}=0$, we find that a non-zero $A_{ee}$ can be obtained even if the Dirac phase $\delta$ is switched off, and its sign is positively correlated to that of $\sin\alpha_{21}$. For $(M_\nu)_{ee}=(M_\nu)_{e\mu}=0$, a large values of $A_{ee}$ is predicted, while $A_{e\tau}$ is always zero.

\begin{figure}
\includegraphics[width=7cm]{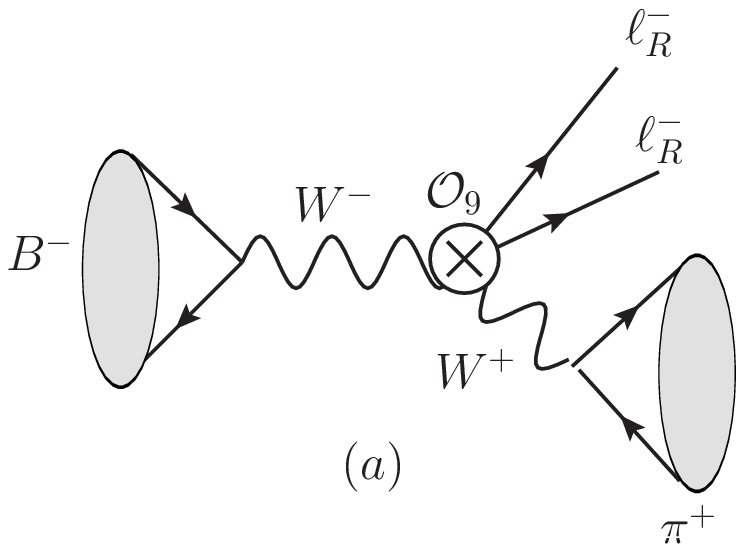}
\includegraphics[width=9cm]{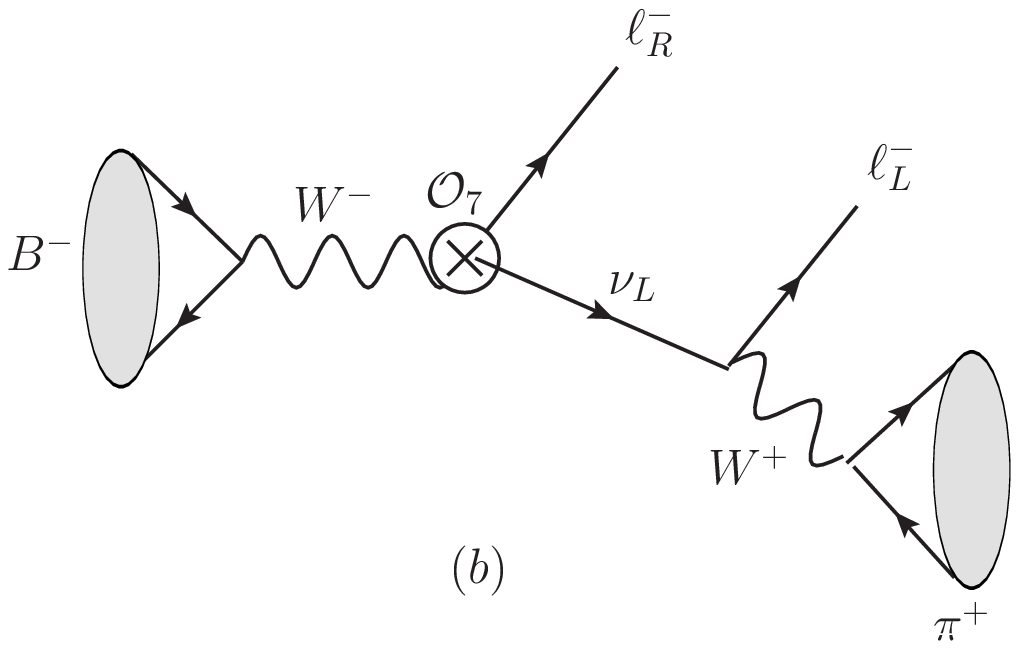}
\caption{Leading-order Feynman diagrams for rare LNV B meson decays induced by (a) ${\cal O}_9$ and (b) ${\cal O}_7$.}
% are positive and negative, respectively.}
\label{BDecay}
\end{figure}

 It is interesting to consider other probes to the Majorana character of the neutrino masses, such as rare LNV meson decays. It is well-known that ordinary channels with Majorana neutrino mass insertions are too small to be observed in the near future. However, it is remarkable that the effective operators, 
such as ${\cal O}_7$ and ${\cal O}_9$,  would give new leading-order contributions. For concreteness, let us consider the process $B^{+} \to \pi^- \mu^+ \mu^+$. If Majorana neutrino masses are induced by  ${\cal O}_9$, the dominant channel to this process is given by the Feynman diagram in Fig.~\ref{BDecay}a, as this tree-level diagram does not involve the tiny Majorana neutrino masses which arise at two-loop level via ${\cal O}_9$. However, with the model parameters fixed by the observed neutrino masses as in Ref.~\cite{Geng:2014gua}, a simple estimation shows that the typical branching ratio for this process is to be of ${\cal O}(10^{-25})$. Other LNV rare meson decays, like $K^+ \to \pi^- \mu^+ \mu^+$, would have even smaller branching ratios. Similar results can also be obtained for ${\cal O}_7$
from Fig.~\ref{BDecay}b. As a result, it seems also impossible to measure such LNV meson decays practically.

\begin{acknowledgments}
The work was supported in part by National Center for Theoretical Science, National Science
Council (NSC-101-2112-M-007-006-MY3), MoST (MoST-104-2112-M-007-003-MY3) and National Tsing Hua
University (104N2724E1).
\end{acknowledgments}

\end{document}